\documentclass[11pt]{article}
\usepackage{amsfonts}

\usepackage{amsmath}
\usepackage{amsmath,amsfonts,amssymb,latexsym}

\newtheorem{theorem}{Theorem}[section]
\newtheorem{definition}{Definition}[section]
\newtheorem{lemma}[theorem]{Lemma}

\newtheorem{notation}[theorem]{Notation}

\newtheorem{corollary}[theorem]{Corollary}
\newtheorem{remark}[theorem]{Remark}
\newenvironment{proof}[1][Proof]{\textsc{#1.} }{\ \rule{0.5em}{0.5em}}
\numberwithin{equation}{section}
\input{tcilatex}

\begin{document}

\title{{\Huge Constraints and Evolution in Cosmology}}
\author{{\Large Yvonne Choquet-Bruhat} \\
University of Paris VI\\
LPTL Tour 22-12\\
4 Place Jussieu, 75252 Paris\\
France\\
\texttt{email: ycb@jussieu.fr} \and {\Large James W. York} \\
Department of Physics\\
Cornell University\\
109 Clark Hall, Ithaca\\
New York 14853-2501, USA\\
\texttt{email: york@mail.lns.cornell.edu}}
\maketitle

\begin{abstract}
We review some old and new results  about strict and non strict hyperbolic
formulations of the classical Einstein equations.
\end{abstract}

\section{Introduction}

The cosmos of general relativity is a pseudo-Riemannian manifold $(\mathcal{V%
},g)$ of Lorentzian signature $(-,+,\ldots ,+)$. The Einstein equations link
its Ricci tensor with a phenomenological tensor which describes the stresses
and energy of the sources. They read: 
\begin{equation*}
\mathrm{Ricci}(g)=\rho ,
\end{equation*}
that is in local coordinates $x^{\lambda},\lambda=0,1,2,\ldots ,n,$ where $%
g=g_{\lambda\mu}dx^{\lambda}dx^{\mu}$ (classical physics $n=3$), 
\begin{equation}
R_{\alpha\beta}=\frac{\partial}{\partial x^{\lambda}}\Gamma_{\alpha\beta}^{%
\lambda}-\frac{\partial}{\partial x^{\alpha}}\Gamma_{\beta\lambda}^{%
\lambda}+\Gamma_{\alpha\beta}^{\lambda}\Gamma
_{\lambda\mu}^{\mu}-\Gamma_{\alpha\mu}^{\lambda}\Gamma_{\beta\lambda}^{\mu
}=\rho_{\alpha\beta} ,  \label{ricciyc}
\end{equation}
where the $\Gamma$'s are the Christoffel symbols: 
\begin{equation*}
\Gamma_{\alpha\beta}^{\lambda}=\frac{1}{2}g^{\lambda\mu}(\frac{\partial }{%
\partial x^{\alpha}}g_{\beta\mu}+\frac{\partial}{\partial x^{\beta}}%
g_{\alpha\mu}-\frac{\partial}{\partial x^{\mu}}g_{\alpha\beta}) ,
\end{equation*}
and $\rho$ is a symmetric 2-tensor given in terms of the stress energy
tensor $T$ by, 
\begin{equation*}
\rho_{\alpha\beta}\equiv T_{\alpha\beta}-\frac{1}{2}g_{\alpha\beta}\,\mathrm{%
tr}\,T,\text{ \ \ with \ \ }\mathrm{tr}\, T\equiv
g^{\lambda\mu}T_{\lambda\mu}.
\end{equation*}

Due to the Bianchi identities the left hand side of the Einstein equations
satisfies the identities, with $\nabla _{\alpha }$ the covariant derivative
in the metric $g$, 
\begin{equation*}
\nabla _{\alpha }(R^{\alpha \beta }-\frac{1}{2}g^{\alpha \beta }\text{ }%
R)\equiv 0,\quad R\equiv g^{\lambda \mu }R_{\lambda \mu }
\end{equation*}
The stress energy tensor of the sources satisfies the conservation laws
which make the equations compatible, 
\begin{equation*}
\nabla _{\alpha }T^{\alpha \beta }=0.
\end{equation*}
In vacuum the stress energy tensor is identically zero. The presence of
sources brings up new problems specific to various types of sources.

Since very little is known about the global properties of the universe, it
is legitimate to study arbitrary manifolds and metrics. Also, in modern
attempts to unify all the fundamental interactions, manifolds of dimension $%
N $ greater than four, endowed with metrics with ($N-4)$-dimensional
isometry groups are considered. We will therefore not restrict our study to
four-dimensional manifolds, when possible.

Since we will treat only non-quantum fields, it seems that a first problem
to look at is the problem of classical dynamics, i.e., the problem of
evolution of initial data. The Einstein equations are a geometric system,
invariant by diffeomorphisms of $\mathcal{V},$ the associated isometries of $%
g,$ and transformation of the sources. From the analyst's point of view they
constitute, for the metric, a system of second order quasi-linear partial
differential equations. The system is determined because the characteristic
determinant is identically zero (a property linked with diffeomorphism
invariance). It is overdetermined because the Cauchy data are not arbitrary
(which must be interpreted geometrically).

To study the Cauchy problem we must split space and time. The evolution will
be formulated for time dependent space tensors.

\section{Moving frame formulas}

\subsection{Frame and coframe}

A moving frame in a subset $\mathcal{U}$ of a differentiable $(n+1)$%
-dimensional manifold $\mathcal{V}$ is a set of $(n+1)$ vector fields on $%
\mathcal{U}$ linearly independent in the tangent space $T_{x}\mathcal{V}$ at
each point $x\in \mathcal{U}.$ A coframe on $\mathcal{U}$ is a set of $(n+1)$
1-forms $\theta ^{\alpha }$ linearly independent at each $x\in \mathcal{U} $
in the dual $T_{x}^{\ast}\mathcal{V}$. In the domain $\mathcal{U}$ of a
chart a coframe is defined by $(n+1)$ linearly independent differential
1-forms, 
\begin{equation}
\theta ^{\alpha }\equiv a_{\beta }^{\alpha }dx^{\beta },
\end{equation}
with $a_{\beta }^{\alpha }$ functions on $\mathcal{U}.$

The metric is Lorentzian if the quadratic form is of Lorentzian signature.

\begin{remark}
The splitting $\mathcal{V}=\mathcal{M}\times \mathbb{R}$, with $\mathcal{M}$
an orientable 3-manifold, implies the existence of a global coframe (but not
of global coordinates!).
\end{remark}

The coframe defined by the 1-forms $\theta^{\alpha}$ is called the natural
frame if $\theta^{\alpha}\equiv dx^{\alpha}.$

In a general frame the differentials of the 1-forms $\theta ^{\alpha }$ do
not vanish; they are given by the 2-forms, 
\begin{equation}
d\theta ^{\alpha }\equiv -\frac{1}{2}C_{\beta \gamma }^{\alpha }\theta
^{\beta }\wedge \theta ^{\gamma }.  \label{2.2}
\end{equation}
The functions $C_{\beta \gamma }^{\alpha }$ on $\mathcal{U}$ are called the
structure coefficients of the frame.

The Pfaff derivative $\partial _{\alpha }$ of a function on $\mathcal{U}$ is
such that, 
\begin{equation}
df\equiv \frac{\partial f}{\partial x^{\alpha }}dx^{\alpha }\equiv \partial
_{\alpha }f\theta ^{\alpha }.
\end{equation}
We denote by $A$ with elements $A_{\alpha }^{\beta }$ of the matrix inverse
of $a$ with elements $a_{\alpha }^{\beta }.$ It holds that: 
\begin{equation}
\partial _{\alpha }f\equiv A_{\alpha }^{\beta }\frac{\partial f}{\partial
x^{\beta }}.
\end{equation}
Pfaff derivatives do not commute. One deduces from (\ref{2.2}) and the
identity $d^{2}f\equiv 0$ that, 
\begin{equation}
d^{2}f\equiv \frac{1}{2}\{\partial _{\beta }\partial _{\gamma }f-\partial
_{\gamma }\partial _{\beta }f-C_{\beta \gamma }^{\alpha }\partial _{\alpha
}f\}\theta ^{\beta }\wedge \theta ^{\gamma }\equiv 0,
\end{equation}
hence, 
\begin{equation}
(\partial _{\alpha }\partial _{\beta }-\partial _{\beta }\partial _{\alpha
})f\equiv C_{\alpha \beta }^{\gamma }\partial _{\gamma }f.
\end{equation}

\subsection{Metric}

A metric on $\mathcal{U}$ is a nondegenerate quadratic form of the $\theta
^{\alpha} $'s: 
\begin{equation}
g\equiv g_{\alpha \beta }\theta ^{\alpha }\theta ^{\beta }.
\end{equation}
A frame is called orthonormal for the metric $g$ if $g_{\alpha \beta }=\pm
1. $ In the case of a Lorentzian metric we will denote by $\theta ^{0}$ the
timelike (co)axis and $\theta ^{i}$ the space (co)axis, then in an
orthonormal frame, $g_{00}=-1$ and $g_{ij}=\delta _{ij},$ the Euclidean
metric.

\subsection{Connection}

A linear connection on $\mathcal{V}$ permits the definition of an intrinsic
derivation of vectors and tensors. It is defined in the domain $\mathcal{U}$
by a matrix-valued 1-form $\omega$ i.e., by a set of matrices $%
\omega_{\gamma}^{\beta}$ linked to functions $\omega_{\alpha\gamma}^{\beta}$
by the identities, 
\begin{equation}
\omega_{\gamma}^{\beta}\equiv\omega_{\alpha\gamma}^{\beta}\theta^{\alpha}.
\end{equation}
The covariant derivative of a vector $v$ with components $v^{\alpha}$ is, 
\begin{equation}
\nabla_{\alpha}v^{\beta}\equiv\partial_{\alpha}v^{\beta}+\omega_{\alpha
\gamma }^{\beta}v^{\gamma}.
\end{equation}
An analogous formula holds for a covariant vector now with a minus sign in
front of $\omega_{\alpha\gamma}^{\beta}$.

\begin{definition}
The connection $\omega$ is called the Riemannian connection of $g$ if

\begin{itemize}
\item  It has vanishing torsion, i.e., 
\begin{equation}
d\theta^{\gamma}+\omega_{\alpha\beta}^{\gamma}\theta^{\alpha}\wedge
\theta^{\beta}=0,
\end{equation}
that is\footnote{%
The interpretation of this condition is that the second covariant
derivatives of scalar functions commute, namely 
\begin{equation*}
\nabla_{\alpha}\partial_{\beta}f-\nabla_{\beta}\partial_{\alpha}f\equiv0.
\end{equation*}
In particular in the natural frame $\omega_{\beta\gamma}^{\alpha}$ is
symmetric in $\beta$ and $\gamma.$}, 
\begin{equation*}
\omega_{\beta\gamma}^{\alpha}-\omega_{\gamma\beta}^{\alpha}=C_{\beta\gamma
}^{\alpha}.
\end{equation*}

\item  The covariant derivative of the metric is zero, i.e., 
\begin{equation}
\partial_{\alpha}g_{\beta\gamma}-\omega_{\alpha\gamma}^{\lambda}g_{\beta%
\lambda}-\omega_{\alpha\beta}^{\lambda}g_{\lambda\gamma}=0.
\end{equation}
\end{itemize}
\end{definition}

These two conditions imply by straightforward computation that, 
\begin{equation}
\omega_{\alpha\gamma}^{\beta}\equiv\Gamma_{\alpha\gamma}^{\beta}+g^{\beta\mu
}\tilde{\omega}_{\alpha\gamma,\mu},  \label{2.12}
\end{equation}
with 
\begin{eqnarray}
\tilde{\omega}_{\alpha\gamma,\mu}&\equiv&\frac{1}{2}(g_{\mu\lambda}C_{\alpha%
\gamma}^{\lambda}-g_{\lambda\gamma}C_{\alpha\mu}^{\lambda}-g_{\alpha%
\lambda}C_{\gamma\mu}^{\lambda}),  \label{2.13} \\
\Gamma_{\alpha\gamma}^{\beta}&\equiv&\frac{1}{2}g^{\beta\mu}(\partial_{%
\alpha
}g_{\gamma\mu}+\partial_{\gamma}g_{\alpha\mu}-\partial_{\mu}g_{\alpha%
\gamma}).  \label{2.14} \\
\notag
\end{eqnarray}
The quantities $\Gamma$ are called the Christoffel symbols of the metric $g.$
The connection coefficients reduce to them in the natural frame. They are
zero for an orthonormal frame.

\subsection{Curvature}

\subsubsection{Definition}

The non-commutativity of covariant derivatives is a geometric property of
the metric. It signals its curvature. The Riemann curvature tensor is
defined as an exterior 2-form with value a linear map in the tangent plane
to $\mathcal{V}$ by the following identity: 
\begin{equation}
(\nabla_{\lambda}\nabla_{\mu}-\nabla_{\mu}\nabla_{\lambda})v^{\alpha}\equiv
R_{\lambda\mu,}{}^{\alpha}{}_{\beta}v^{\beta},
\end{equation}
which gives by straightforward identification, 
\begin{equation}
R_{\lambda\mu,}{}^{\alpha}{}_{\beta}\equiv\partial_{\lambda}\omega_{\mu\beta
}^{\alpha}-\partial_{\mu}\omega_{\lambda\beta}^{\alpha}+\omega_{\lambda\rho
}^{\alpha}\omega_{\mu\beta}^{\rho}-\omega_{\mu\rho}^{\alpha}\omega
_{\lambda\beta}^{\rho}-\omega_{\rho\beta}^{\alpha}C_{\lambda\mu}^{\rho}.
\end{equation}

\subsubsection{Symmetries and antisymmetries}

The Riemann tensor is a symmetric double 2-form: it is antisymmetric in its
first two indices, and in its last two indices written in covariant form. It
is invariant by the interchange of these two pairs.

\subsubsection{Bianchi identities}

It holds that (vanishing of the covariant differential of the curvature
2-form): 
\begin{equation}
\nabla _{\alpha }R_{\beta \gamma ,\lambda \mu }+\nabla _{\beta }R_{\gamma
\alpha ,\lambda \mu }+\nabla _{\gamma }R_{\alpha \beta ,\lambda \mu }\equiv
0.
\end{equation}

\subsubsection{Ricci tensor, scalar curvature, Einstein tensor}

The Ricci tensor is defined by 
\begin{equation}
R_{\alpha \beta }\equiv R_{\lambda \alpha ,}{}^{\lambda }{}_{\beta }.
\end{equation}
The scalar curvature is, 
\begin{equation}
R\equiv g^{\alpha \beta }R_{\alpha \beta }.
\end{equation}
The Einstein tensor is, 
\begin{equation}
S_{\alpha \beta }\equiv R_{\alpha \beta }-\frac{1}{2}g_{\alpha \beta }R.
\end{equation}

\subsubsection{Conservation identity}

Contracting the Bianchi identities gives that, 
\begin{equation}
\nabla _{\alpha }R_{\beta \gamma ,}{}^{\alpha }{}_{\mu }-\nabla _{\beta
}R_{\gamma \mu }+\nabla _{\gamma }R_{\beta \mu }\equiv 0,
\end{equation}
and a further contraction gives the following identity satisfied by the
Einstein tensor: 
\begin{equation}
\nabla _{\alpha }S^{\alpha \beta }\equiv 0.
\end{equation}
This identity implies that the sources must satisfy the so-called
conservation laws, 
\begin{equation}
\nabla _{\alpha }T^{\alpha \beta }=0.
\end{equation}

\section{(n+1)-splitting adapted to space slices}

\label{3}

\subsection{Definitions}

We consider a spacetime with manifold $\mathcal{V}=\mathcal{M}\times\mathbb{%
\ R} $ and hyperbolic metric $g$ such that the submanifolds $\mathcal{M}%
_{t}\equiv \mathcal{M}\times \{t\}$ are spacelike. We take a frame with
space axis $e_{i}$ tangent to the space slice $\mathcal{M}_{t}$ and time
axis $e_{0}$ orthogonal to it. Such a frame is particularly adapted to the
solution of the Cauchy problem and will be called \textbf{a Cauchy adapted
frame}. The dual coframe is: 
\begin{equation}
{\theta }^{i}=dx^{i}+\beta ^{i}dt,
\end{equation}
$\beta ^{i}$ is a time dependent vector tangent to $\mathcal{M}_{t}$ called
the shift. The 1-form $\theta ^{0}$ does not contain $dx^{i}.$ We choose: 
\begin{equation}
\theta ^{0}=dt.
\end{equation}
The pfaffian derivatives (action of the vector basis $e_{\alpha})$ with
respect to the adapted coframe are, 
\begin{equation}
\partial_{0}={\partial}_{t}-{\beta}^{j}{\partial}_{j},\quad {\partial }_{i}={%
\partial}/{\partial}x^{i},\text{ \ \ with \ \ }{\partial}_{t}={\ \partial}/{%
\partial}t .  \label{vectoryc}
\end{equation}
In the coframe $\theta ^{\alpha }$ the metric reads, 
\begin{equation}
ds^{2}=-N^{2}({\theta }^{0})^{2}+g_{ij}{\theta }^{i}{\theta }^{j}.
\end{equation}
The function $N$ is called the lapse. We shall assume throughout $N>0$ and
the space metric $\bar{g}$ induced by $g$ on $\mathcal{M}_{t}$ properly
Riemannian. An overbar will denote a spatial tensor or operator, i.e., a $t$%
-dependent tensor or operator on $\mathcal{M}$. Note that $\bar{g}%
_{ij}=g_{ij}$ and $\bar{g} ^{ij}=g^{ij}$.

\subsection{Structure coefficients}

The structure coefficients of a frame of a Cauchy adapted frame are found to
be, 
\begin{equation}
C_{0j}^{i}=-C_{j0}^{i}={\partial }_{j}{\beta }^{i},
\end{equation}
and all other structure coefficients are zero.

\subsection{Splitting of the connection}

We denote by $\bar{\nabla}$ covariant derivatives in the space metric $\bar{%
g }.$ Using the general formulas (\ref{2.12}), (\ref{2.13}), (\ref{2.14}) we
find that, 
\begin{equation}
{\omega }^{i}{}_{jk}={\Gamma }^{i}{}_{jk}=\bar{\Gamma}_{jk}^{i},
\end{equation}
\begin{equation}
{\omega }^{i}{}_{00}=Ng^{ij}{\partial }_{j}N,\text{ \ \ }{\omega }
^{0}{}_{0i}={\omega }^{0}{}_{i0}=N^{-1}{\partial }_{i}N,\text{ \ \ }{\omega }
^{0}{}_{00}=N^{-1}{\partial }_{0}N,  \notag
\end{equation}
and 
\begin{equation}
{\omega }^{0}{}_{ij}={\frac{1}{2}}N^{-2}\{\partial {_{0}}
g_{ij}+g_{hj}C_{i0}^{h}+g_{ih}C_{j0}^{h}),
\end{equation}
that is, 
\begin{equation}
{\omega }^{0}{}_{ij}={\frac{1}{2}}N^{-2}\{\partial {_{0}}g_{ij}-g_{hj}%
\partial _{i}\beta ^{h}-g_{ih}\partial _{j}\beta ^{h}).
\end{equation}
Using the expression (\ref{vectoryc}) of $\partial _{0}$ we obtain that, 
\begin{equation}
{\omega }^{0}{}_{ij}={\frac{1}{2}}N^{-2}{\hat{\partial}_{0}}g_{ij},
\end{equation}
where the operator $\hat{\partial}_{0}$ is defined on any $t$-dependent
space tensor $\mathbf{T}$ by the formula, 
\begin{equation}
{\hat{\partial}_{0}}={\frac{{\partial }}{{\partial }t}}-\bar{L}_{{\beta }},
\end{equation}
where $\bar{L}_{{\beta }}$ is the Lie derivative on $\mathcal{M}_{t}$ with
respect to the spatial vector $\mathbf{\beta }$. Note that ${\hat{\partial}%
_{0}}\mathbf{\ T}$ is a $t$-dependent space tensor of the same type as $%
\mathbf{T.}$

The extrinsic curvature $K_{ij}$ (second fundamental tensor) of $\mathcal{M}%
_{t}$ is classically defined as the projection on $\mathcal{M}_{t}$ of the
covariant derivative of the unit normal $\nu $, past-oriented, that is 
\begin{equation}
K_{ij}\equiv \frac{1}{2}(\nabla _{i}\nu _{j}+\nabla _{j}\nu _{i}),
\end{equation}
with, in our coframe, 
\begin{equation}
\nu _{i}=0,\text{ \ \ }\nu _{0}=N.
\end{equation}

\begin{lemma}
\label{Lemma 1} The following identity holds, 
\begin{equation}
\hat{\partial}_{0}g_{ij}\equiv -2NK_{ij}.
\end{equation}
\end{lemma}

\begin{proof}
It holds that, 
\begin{equation}
\nabla_{i}\nu_{j}=-\omega_{ij}^{0}\nu_{0}=-{\frac{1}{2}}N^{-1}{\hat{\partial 
}_{0}}g_{ij}=K_{ij}.
\end{equation}
The remaining connection coefficients are found to be, 
\begin{equation}
{\omega}^{i}{}_{j0}=-NK^{i}{}_{j}+{\partial}_{j}{\beta}^{i},\text{ \ \ \ }{\
\omega}^{i}{}_{0j}=-NK^{i}{}_{j},
\end{equation}
and this completes the proof of the Lemma.
\end{proof}

\subsection{Splitting of the Riemann tensor}

We deduce from the general formula giving the Riemann tensor and the
splitting of the connection the following identities, 
\begin{equation}
R_{ij,kl}=\bar{R}_{ij,kl}+K_{ik}K_{lj}-K_{il}K_{kj},  \label{3.16}
\end{equation}
\begin{equation}
R_{0i,jk}=N{(\bar{\nabla}}_{j}K_{ki}-{\bar{\nabla}}_{k}K_{ji}),  \label{3.17}
\end{equation}
\begin{equation}
R_{0i,0j}=N({\hat{\partial}_{0}}K_{ij}+NK_{ik}K^{k}{}_{j}+{\bar{\nabla}}_{i}{%
\partial }_{j}N).
\end{equation}
From these formulae one obtains the following ones for the Ricci curvature: 
\begin{equation}
NR_{ij}=N\bar{R}_{ij}-{\hat{\partial}_{0}}%
K_{ij}+NK_{ij}K_{h}^{h}-2NK_{ik}K^{k}{}_{j}-{\bar{\nabla}}_{i}{\partial }%
_{j}N,  \label{3.20}
\end{equation}
\begin{equation}
N^{-1}R_{0j}={\partial }_{j}K_{h}^{h}-{\bar{\nabla}}_{h}K^{h}{}_{j},
\end{equation}
\begin{equation}
R_{00}=N({\hat{\partial}_{0}}K_{h}^{h}-NK_{ij}K^{ij}+\bar{\Delta}N).
\end{equation}
Also, 
\begin{equation}
g^{ij}R_{ij}=\bar{R}-N^{-1}{\hat{\partial}_{0}}%
K_{h}^{h}+(K_{h}^{h})^{2}-N^{-1}\bar{\Delta}N,
\end{equation}
\begin{equation}
S_{00}\equiv R_{00}-{\frac{1}{2}}g_{00}R\equiv {\frac{1}{2}}%
(R_{00}+g^{ij}R_{ij}),  \label{3.24}
\end{equation}
hence, 
\begin{equation}
2N^{-2}S_{00}\equiv -2S_{0}^{0}\equiv \bar{R}-K_{ij}K^{ij}+(K_{h}^{h})^{2},
\end{equation}
with $\bar{R}=g^{ij}\bar{R}_{ij}$.

\section{Constraints and evolution}

We see in the above decomposition of the Ricci tensor that none of the
Einstein equations contains the time derivatives of the lapse $N$ and shift $%
\beta .$ One is thus led to consider the Einstein equations as a dynamical
system for the two fundamental forms $\bar{g}$ and $K$ of the space slices $%
\mathcal{M}_{t}.$ This dynamical system splits as follows.

\textbf{Constraints}

The restriction to $\mathcal{M}_{t}$ of the right hand side of the
identities (\ref{3.20}) and (\ref{3.24}) contains only the metric $g_{ij}$
and the extrinsic curvature $K_{ij}$ of $\mathcal{M}_{t}$ as tensor fields
on $\mathcal{M}_{t}.$ When the Einstein equations are satisfied, i.e., when, 
\begin{equation}
S_{\alpha \beta }\equiv R_{\alpha \beta }-\frac{1}{2}g_{\alpha \beta }\
R=T_{\alpha \beta }\equiv \rho _{\alpha \beta }-\frac{1}{2}g_{\alpha \beta
}\rho ,
\end{equation}
these identities lead to the following equations called constraints:

Momentum constraint

\begin{equation}
C_{i}\equiv \frac{1}{N}(R_{0i}-\rho _{0i})\equiv -\bar{\nabla}_{h}K_{i}^{h}+ 
\bar{\nabla}_{i}K_{h}^{h}-N^{-1}\rho _{0i},  \label{4.2}
\end{equation}

Hamiltonian constraint

\begin{equation}
C_{0}\equiv \frac{2}{N^{2}}(S_{00}-T_{00})\equiv \bar{R}
-K_{j}^{i}K_{i}^{j}+(K_{h}^{h})^{2}+2T_{0}^{0}.  \label{4.3}
\end{equation}

These constraints are transformed into a system of elliptic equations on
each submanifold $\mathcal{M}_{t},$ in particular on $\mathcal{M}_{0}$, for $%
\overset{\_}{ g}=g_{0},K=K_{0},$ by the conformal method (cf. C-B and York
1980, Isenberg 1995, C-B, Isenberg and York 2000).

\textbf{Evolution}

The equations, 
\begin{equation*}
R_{ij}\equiv \bar{R}_{ij}-\frac{\hat{\partial}_{0}K_{ij}}{N}%
-2K_{jh}K_{i}^{h}+K_{ij}K_{h}^{h}-\frac{\bar{\nabla}_{j}\partial _{i}N}{N}%
=\rho _{ij},
\end{equation*}
together with the definition, 
\begin{equation*}
\hat{\partial}_{0}g_{ij}=-2NK_{ij},
\end{equation*}
determine the derivatives transversal to $\mathcal{M}_{t}$ of $\bar{g}$ and $%
K$ when these tensors are known on $\mathcal{M}_{t}$ as well as the lapse $N$
and shift $\beta ,$ and source $\rho _{ij}.$ It is natural to look at these
equations as evolution equations determining $\bar{g}$ and $K$, while $N$
and $\beta ,$ projections of the tangent to the time line respectively on $%
e_{0}$ and the tangent space to $\mathcal{M},$ are considered as gauge
variables. This point of view is supported by the following theorem
(Anderson and York 1997, previously given for sources in C-B and
Noutchegueme 1988):

\begin{theorem}
When $R_{ij}-\rho_{ij}=0$ the constraints satisfy a linear homogeneous first
order symmetric hyperbolic system, they are satisfied if satisfied initially.
\end{theorem}

\begin{proof}
When $R_{ij}-\rho _{ij}=0$ we have, in the Cauchy adapted frame, with $\rho
\equiv g^{\alpha \beta }\rho _{\alpha \beta ,}$%
\begin{equation*}
R-\rho =-N^{2}(R^{00}-\rho ^{00}),
\end{equation*}
hence, 
\begin{equation*}
S^{00}-T^{00}=\frac{1}{2}(R^{00}-\rho ^{00})\text{ \ \ and \ }R-\rho
=-2N^{2}(S^{00}-T^{00})=2(S_{0}^{0}-T_{0}^{0}),
\end{equation*}
and 
\begin{equation*}
S^{ij}-T^{ij}=-\frac{1}{2}\bar{g}^{ij}(R-\rho )=-\bar{g}%
^{ij}(S_{0}^{0}-T_{0}^{0}),
\end{equation*}
the Bianchi identities give therefore a linear homogeneous system for $%
\Sigma _{0}^{i}\equiv S_{0}^{i}-T_{0}^{i}$ and $\Sigma _{0}^{0}\equiv
S_{0}^{0}-T_{0}^{0}$ with principal parts, 
\begin{equation*}
N^{-2}\partial _{0}\Sigma _{0}^{i}+\bar{g}^{ij}\partial _{j}\Sigma _{0}^{0},%
\text{ \ and \ \ }\partial _{0}\Sigma _{0}^{0}+\partial _{i}\Sigma _{0}^{i}.
\end{equation*}
This system is symmetrizable hyperbolic, it has a unique solution, zero if
the initial values are zero. The characteristic which determines the domain
of dependence is the light cone.
\end{proof}

\section{Analytic Cauchy problem}

Geometrical initial data for the Einstein equations are a triple $(\mathcal{M%
},\bar{g}_{0},K_{0})$ with $\bar{g}_{0}$ a properly Riemannian metric on the
n-dimensional manifold $\mathcal{M}$, and $K_{0}$ a symmetric 2-tensor on $%
\mathcal{M}$. A solution of the Cauchy problem for the Einstein equations
with these initial data is an (n+1)-dimensional pseudo-Riemannian manifold $(%
\mathcal{V},g)$ which we shall suppose of signature $(-,+,\dots ,+)$, such
that $\mathcal{M}$ can be identified with a submanifolds $\mathcal{M}_{0}$
of $\mathcal{V}$, with $\bar{g}_{0}$ the metric induced by $g$ on $\mathcal{M%
}_{0}$ and $K_{0}$ the extrinsic curvature of $\mathcal{M}_{0}$ as
submanifold of $(\mathcal{V},g)$. The manifold $(\mathcal{V},g)$ is called
an \textit{Einsteinian development} of the data. When the sources are given,
for example zero (vacuum case), these geometrical initial data cannot be
chosen arbitrarily, they must satisfy on $\mathcal{M}$ the constraints (\ref
{4.2}), (\ref{4.3}).

The evolution equations read: 
\begin{eqnarray}
\partial _{t}g_{ij}&=&-2NK_{ij}+\bar{\nabla}_{i}\beta _{j}+\bar{\nabla}
_{j}\beta _{i}, \\
\partial _{t}k_{ij}&=&N\{\bar{R}_{ij}-2K_{ih}K_{j}^{h}+K_{ij}K_{h}^{h}\}-%
\bar{ \nabla}_{i}\partial _{j}N  \notag \\
&+& \beta ^{h}\bar{\nabla}K_{ij}+k_{ih}\bar{\nabla}_{j}\beta ^{h}+k_{hj}\bar{%
\nabla}_{i}\beta ^{h}-N\rho _{ij}.
\end{eqnarray}
No equation contains the time derivatives of the lapse $N$ and shift $\beta$%
. We suppose that these quantities are given on $\mathcal{V}.$ The system is
of the Cauchy-Kovalevski type, therefore we have the following theorem:

\begin{theorem}
If the initial data are analytic on $\mathcal{M}_{0}$ while the sources, the
shift and the lapse are analytic in a neighborhood of $\mathcal{M}_{0}$ then
there exists a neighborhood of $\mathcal{M}_{0}$ in $\mathcal{M}\times 
\mathbb{R}$ such that the evolution equations have a solution in this
neighborhood taking these Cauchy data.
\end{theorem}

We deduce from this Theorem the following one:

\begin{theorem}
If the sources satisfy the conservation laws, in particular are zero, on $%
\mathcal{V}$ the solution of the evolution equations satisfies the full
Einstein equations if the initial data satisfy the constraints.
\end{theorem}

Since the conservation laws depend also on the metric the application of
these theorems requires further study, except in the case where the sources
are zero. In this vacuum case we can enunciate:

\begin{corollary}
Analytic initial data satisfying the vacuum constraints admit a vacuum
Einsteinian development.
\end{corollary}

\begin{proof}
Take an arbitrary analytic shift $\beta $ and $N>0.$
\end{proof}

We will come back later to the geometric uniqueness problem.

\section{Non-strict hyperbolicity of $R_{ij}=0$}

An evolution part of Einstein equations should exhibit causal propagation,
i.e., with domain of dependence determined by the light cone of the
spacetime metric.

The equations $R_{ij}=0$ are, when $N$ and $\beta $ are known, a second
order differential system for $g_{ij}.$ The hyperbolicity of a quasi-linear
system is defined through the linear differential operator obtained by
replacing in the coefficients the unknown by given values. In our case and
for given $N,\beta $ and $g_{ij}$, the principal part of this operator
acting on a symmetric 2-tensor $\gamma _{ij}$ is, 
\begin{equation*}
\frac{1}{2}\{(N^{-2}\partial _{00}^{2}-g^{hk}\partial _{hk}^{2})\gamma
_{ij}+\partial ^{k}\partial _{j}\gamma _{ik}+\partial ^{k}\partial
_{i}\gamma _{jk}-g^{hk}\partial _{i}\partial _{j}\gamma _{hk}\}.
\end{equation*}
The characteristic matrix $\mathcal{M}$ at a point of spacetime is the
linear operator obtained by replacing the derivation $\partial $ by a
covariant vector $\xi .$ The characteristic determinant is the determinant
of this linear operator. For simplicity we compute it in the classical case
of space dimension $n=3.$ We find, 
\begin{equation*}
Det\mathcal{M}=\xi _{0}^{6}(g^{\alpha \beta }\xi _{\alpha }\xi _{\beta
}\}^{3}.
\end{equation*}
The characteristic cone is the dual of the cone defined in the cotangent
plane by annulation of the characteristic polynomial. For our system the
characteristic cone splits into the light cone of the given spacetime metric
and the normal to its space slice. Since these characteristics appear as
multiple and the system is non-diagonal, it is not hyperbolic in the usual
sense. One can prove by diagonalization of the system the following theorem
(C-B 2000).

\begin{theorem}
When $N>0$ and $\beta$ are given, arbitrary, the system $R_{ij}=0$ is a
non-strict hyperbolic system in the sense of Leray-Ohya for $g_{ij}$, in the
Gevrey class $\gamma=2,$ as long as $g_{ij}$ is properly Riemannian. If the
Cauchy data as well as $N$ and $\beta$ are in such a Gevrey class, the
Cauchy problem has a local in time solution with domain of dependence
determined by the light cone.
\end{theorem}

\section{Wave equation for $K,$ Hyperbolic system}

Various hyperbolic systems have been obtained in recent years for the
evolution of the dynamical variables $(g_{ij},K_{ij})$ by linear combination
of $R_{ij}$ with the constraints. The first of these hyperbolic systems has
been obtained in C.B and Ruggeri 1983, zero shift, extended in C.B and York
1995 to an arbitrary shift. It uses a quasi-diagonal system of wave
equations for $K_{ij},$ shown to hold modulo a gauge condition for $N,$
called now `densitizing the lapse'. It works as follows.

We use the expressions for $R_{0i}$ and $R_{ij},$ together with $\hat{%
\partial }_{0}g_{ik}=-2NK_{ik}$ which imply, 
\begin{equation}
\hat{\partial}_{0}\bar{\Gamma}_{ij}^{h}=-g^{hk}\{\bar{\nabla}_{i}(NK_{jk}+ 
\bar{\nabla}_{j}(NK_{ik})-\bar{\nabla}_{k}(NK_{ij})\},
\end{equation}
to obtain the identity, with $f_{(ij)}=f_{ij}+f_{ji},$ and $H\equiv \text{tr}%
(K)\equiv K_{i}^{i}$, 
\begin{eqnarray}
\Omega_{ij}&\equiv&\hat{\partial}_{0}R_{ij}-\bar{\nabla}_{(i}R_{j)0}\equiv -%
\hat{\partial}_{0}(N^{-1}\hat{\partial}_{0}K_{ij})+\bar{\nabla}_{h}\bar{
\nabla}^{h}(NK_{ij})-\hat{\partial}_{0}(N^{-1}\bar{\nabla} _{j}\partial_{i}N)
\notag \\
&-& N\bar{\nabla}i\partial_{j}H +\hat{\partial}%
_{0}(HK_{ij}-2K_{im}K_{j}^{m})-\bar{\nabla} _{(i}(K_{j)h}\allowbreak%
\partial^{h}N)-2N\bar{R}_{ijm}^{h}K_{h}^{m}  \notag \\
&-& N\bar{R}_{m(i}K_{j)}^{m}+H\bar{\nabla}_{j}\partial_{i}N.
\end{eqnarray}
This identity shows that for a solution of the Einstein equations, 
\begin{equation*}
R_{\alpha\beta}=\rho_{\alpha\beta},
\end{equation*}
the extrinsic curvature $K$ satisfies a second order differential system
which is quasi-diagonal with principal part the wave operator, except for
the terms $-N\bar{\nabla}_{i}\partial_{j}H.$ The unknown $\bar{g}$ appears
at second order, as well as $N$ except for the term $-\hat{\partial}%
_{0}(N^{-1}\bar{\nabla}_{j}\partial_{i}N)$. It holds that, 
\begin{equation}
\hat{\partial}_{0}(N^{-1}\bar{\nabla}_{j}\partial_{i}N)+N\bar{\nabla}
_{j}\partial_{i}H\equiv N^{-1}\bar{\nabla}_{j}\partial_{i}(
\partial_{0}N+N^{2}H)+X_{ij},
\end{equation}
where $X_{ij}$ is only of first order in $\bar{g}$ and $K.$

We then densitize the lapse, that is we set, 
\begin{equation}
N=\alpha (\det \bar{g})^{1/2},
\end{equation}
where $\alpha $ is an arbitrary positive tensor density of weight $-1$
called the `densitized lapse'. Then, using the formula for the derivative of
a determinant and the identity from Lemma \ref{Lemma 1}, we find that: 
\begin{equation}
\partial _{0}N=-N^{2}H+\hat{\partial}_{0}\alpha (\det \bar{g})^{1/2}.
\label{7.5}
\end{equation}
We see that $\partial _{0}N+N^{2}H$ is an algebraic function of $\bar{g}.$
Hence, we have proved the following theorem.

\begin{theorem}
Let $\alpha>0$ and $\beta$ be an arbitrary space tensor density and space
tensor, depending on $t$. Set $N=\alpha(\det\bar{g})^{1/2}$, then:

1. The equations 
\begin{equation}
\Omega_{ij}=0,  \label{7.7}
\end{equation}
are a quasi-diagonal system of wave equations for $K.$

2. The equations above together with 
\begin{equation}
\hat{\partial}_{0}g_{ij}=-2NK_{ij},  \label{7.8}
\end{equation}
are a hyperbolic Leray system for $\bar{g}$ and $K.$
\end{theorem}

\begin{proof}
Part 1 has already been proved. To prove part 2, we give to the equations
and unknown the following weights: 
\begin{equation}
m(K)=2,\, m(\bar{g})=2,\, n(\Omega=0)=0,\, n(\hat{\partial}g=2NK)=1.
\end{equation}
The principal operator is then a matrix diagonal by blocks. Each block,
corresponding to a pair $(ij)$ of indices, is given by,

$\left( 
\begin{tabular}{cc}
$-N^{2}\partial_{00}^{2}+g^{ij}\partial_{i}\partial_{j}$ & $X$ \\ 
$0$ & $\partial_{0}$%
\end{tabular}
\right) $ \ 

with $X$ a second order operator. The characteristic determinant is ($%
-N^{2}\xi _{0}^{2}+g^{ij}\xi _{i}\xi _{j})\xi _{0},$ it is a hyperbolic
polynomial.
\end{proof}

\section{Hyperbolic-elliptic system.}

An alternative method to the densitization of the lapse is to consider $H$
as a given function $h$ on space time, i.e., imposing given mean extrinsic
curvature on the space slices. The second order equation for $K$ obtained
above reduces again to a quasi-diagonal system with principal part the wave
operator. This gauge condition was used by Christodoulou and Klainerman,
with $h$ = 0, in the asymptotically Euclidean case, in the general case by
C.B and York 1996. The lapse $N$ is then determined through the equation $%
R_{0}^{0}=\rho _{0}^{0}$ which now reads in the general (non-vacuum) case, 
\begin{equation*}
\bar{\nabla}^{i}\partial _{i}N-(K_{ij}K^{ij}-\rho _{0}^{0})N=-\partial _{0}h.
\end{equation*}
This equation is an elliptic equation for N when $\bar{g}$ , $K$ and $\rho $
are known.

Note that for energy sources satisfying the energy condition we have $%
-\rho_{0}^{0}\geq0$ as well as $\mid K\mid^{2}\equiv K_{ij}K^{ij}\geq0$, an
important property for the solution of the elliptic equation. The mixed
hyperbolic-elliptic system that we have constructed will determine the
unknowns $N$ and $\bar{g}$ in a neighborhood of $\mathcal{M}$ in $\mathcal{M}%
\times \mathbb{R}$ when the shift $\beta$ is chosen.

\section{Local existence and uniqueness}

Known properties of hyperbolic systems on manifolds and the use of the
Bianchi identities to show the preservation of constraints under evolution,
lead to the following theorem.

\begin{theorem}
\label{theorem 6} Let ($\mathcal{M},\bar{g}_{0},K_{0})$ be an initial data
set satisfying the vacuum constraints, where $\mathcal{M}$ is a smooth $n$%
-dimensional manifold endowed with a smooth, Sobolev regular, Riemannian
metric $e,$ and where $\bar{g}_{0}\in H_{s}^{u.loc}$, $K_{0}\in
H_{s-1}^{u.loc}$ are a properly Riemannian metric and a symmetric 2-tensor
on $\mathcal{M}$ respectively. Suppose arbitrarily given on $\mathcal{M}%
\times \mathcal{I},$ $\mathcal{I}$ an interval of $\mathbb{R},$ the gauge
variables $\alpha ,\beta \in C^{0}(I,H_{s}^{u.loc})\cap
C^{1}(I,H_{s-1}^{u.loc})\cap C^{2}(I,H_{s-2}^{u.loc}).$ Then if $s>\frac{n}{2%
}+1$ there exists an interval $\mathcal{J}\subset\mathcal{\ I}$ and a
Lorentzian metric, 
\begin{equation}
g=-N^{2}dt^{2}+g_{ij}(dx^{i}+\beta ^{i}dt)(dx^{j}+\beta ^{j}dt),\text{ \ \ }%
N=(\alpha \det \bar{g})^{1/2},
\end{equation}
solution of the given Cauchy problem on $\mathcal{M}_{t_{0}},$ $t_{0}\in 
\mathcal{J},$ for the vacuum Einstein equations. For a given pair $\alpha
,\beta $ this solution is unique.
\end{theorem}

\begin{notation}
The spaces $H_{s}^{u.loc}$ are spaces of tensors which have generalized
covariant derivatives in the metric $e$ of order up to $s$ which are square
integrable on each open set of some given covering of $\mathcal{M},$ with
norm uniformly bounded (i.e., for a given tensor, independent of the
subset). The manifold $(\mathcal{M},e)$ is called Sobolev regular if the
covering can be chosen so that these Sobolev spaces satisfy the usual
embedding and multiplication properties. This is always the case if ($%
\mathcal{M},e)$ is complete.
\end{notation}

Two isometric space times $(\mathcal{V},g)$ and $(\tilde{\mathcal{V}},\tilde{%
g})$ are considered as identical. One can prove the following theorem.

\begin{theorem}[Physical uniqueness]
\begin{enumerate}
\item  Let ($\mathcal{M}\times \tilde{\mathcal{J}},\tilde{g})$ be another
solution of the Cauchy problem of Theorem \ref{theorem 6}, with a different
gauge choice. There exists an isometry of ($\mathcal{M}\times \tilde{%
\mathcal{J}}^{\prime },\tilde{g})$ onto ($\mathcal{M}\times \mathcal{J}%
^{\prime },g),$ with $\mathcal{J}^{\prime }\subset \mathcal{J}$ and $\tilde{%
\mathcal{J}}^{\prime }\subset \tilde{\mathcal{J}}.$

\item  The solution of the Cauchy problem for the vacuum Einstein equations
is geometrically unique (i.e., up to isometries) in the class of globally
hyperbolic space times.
\end{enumerate}
\end{theorem}

For the definition of globally hyperbolic spacetimes see S. Cotsakis'
lectures, this volume.

\section{First order hyperbolic systems}

Such systems are supposed to be more amenable to numerical computation.

\subsection{FOSH systems}

A first order system of $N$ equations which reads, 
\begin{equation}
M^{\alpha IJ}(u)\partial _{\alpha }u_{I}+f^{J}(u)=0,  \label{10.1}
\end{equation}
where $u_{I}$, $I=1,\dots ,N$, are a set of unknowns (for instance the
components of one or several tensors), is called symmetric if the matrices $%
M^{\alpha }$ are symmetric. Such a symmetric system is hyperbolic for the
space slices $\mathcal{M}_{t}$ if the matrix $\tilde{M}^{t},$ coefficient of 
$\partial /\partial t$ is positive definite. The energy associated to such
systems is straightforward to write in a Cauchy adapted frame. In this case $%
\tilde{M} ^{t}=M^{0},$ while $\tilde{M}^{i}=M^{i}-\beta ^{i}M^{0}.$
Equations (\ref{10.1}) imply, 
\begin{equation}
u_{J}M^{\alpha IJ}(u)\partial _{\alpha }u_{I}+u_{J}f^{J}(u)\equiv \frac{1}{2}%
\partial _{\alpha }(M^{\alpha IJ}u_{I}u_{J})+F(u)=0.  \label{above}
\end{equation}
We express the $\partial _{\alpha }$'s as linear combinations of the usual
time and space partial derivatives. Then, Eq. (\ref{above}) takes the form, 
\begin{equation}
u_{J}\tilde{M}^{\alpha IJ}(u)\frac{\partial }{\partial x^{\alpha }}
u_{I}+u_{J}f^{J}(u)\equiv \frac{1}{2}\frac{\partial }{\partial x^{\alpha }}( 
\tilde{M}^{\alpha IJ}u_{I}u_{J})+F(u)=0.
\end{equation}
Integrating this equation on a strip $\mathcal{M}\times [0,T]$ leads to an
energy equality and an energy inequality (see later `Bel-Robinson energy').

Anderson, C-B and York 1995 have written the system (\ref{7.7}), (\ref{7.8})
together with the gauge condition (\ref{7.5}) as a FOSH system, using also
the equation $R_{00}=0$.

A FOSH system had also been obtained by Frittelli and Reula 1994 just by
combination of $R_{ij}$ with the constraints and densitization of the lapse.
They used it to discuss the Newtonian approximation. A number of variants
has been written since then, and their quality for numerical computation
discussed (see in particular the Einstein-Christoffel system of Anderson and
York 1999).

\subsection{Other first order hyperbolic systems}

Another criterion than symmetry has been recently used to test hyperbolicity
of first order systems: It is the number of linearly independent
eigenvectors associated to a multiple characteristic. If this number is
equal to the multiplicity, and modulo conditions of uniformity, the system
is hyperbolic. The verification of such a property involves heavy
computations. Kidder, Scheel and Teukolsky 2001 write a whole family of such
systems. They introduce as new unknowns the partial derivatives $\partial
_{h}g_{ij}$. They show that densitization of the lapse is a necessary
condition for the hyperbolicity of the obtained systems. They evolve some of
them numerically in the case of a one-black-hole spacetime and discuss their
accuracy, i.e., how well the constraints are preserved.

\section{Bianchi-Einstein equations}

The Riemann tensor is the geometric object which intrinsically defines
gravitational effects. The following results bring nothing new for the local
existence and unicity theorems but they are useful for the global-in-time
studies.

\subsection{Wave equation for the Riemann tensor}

The Riemann tensor of a pseudo-Riemannian metric, $R_{\alpha \beta ,\lambda
\mu }$, is antisymmetric in its pair of first indices as well as in its pair
of last indices. We call it a symmetric double 2-form because it possesses
the symmetry, 
\begin{equation}
R_{\alpha \beta ,\lambda \mu }\equiv R_{\lambda \mu ,\alpha \beta }.
\label{11.1}
\end{equation}
The Riemann tensor satisfies the Bianchi identities 
\begin{equation}
\nabla _{{\alpha }}R_{{\beta }{\gamma },{\lambda }\mu }+\nabla _{{\gamma }
}R_{{\alpha }{\beta },{\lambda }\mu }+\nabla _{{\beta }}R_{{\gamma }{\alpha }
,{\lambda }\mu }\equiv 0.  \label{11.2}
\end{equation}
These identities imply by contraction, 
\begin{equation}
\nabla _{{\alpha }}R_{{\beta }{\gamma },\dots\mu }^{\dots\alpha }+\nabla _{{%
\ \gamma }}R_{{\alpha }{\beta },\dots\mu }^{\dots\alpha }+\nabla _{{\beta }%
}R_{{\ \gamma }{\alpha },{\dots}\mu }^{\dots\alpha }\equiv 0.
\end{equation}
Using the symmetry (\ref{11.1}) gives the identities: 
\begin{equation}
\nabla _{{\alpha }}R^{{\alpha }}\mathstrut _{{\beta },{\lambda }\mu }+\nabla
_{\mu }R_{{\lambda }{\beta }}-\nabla _{{\lambda }}R_{\mu {\beta }}\equiv 0.
\end{equation}
If the Ricci tensor $R_{\alpha \beta \text{ }}$ satisfies the Einstein
equations, 
\begin{equation}
R_{{\alpha }{\beta }}=\rho _{{\alpha }{\beta }},
\end{equation}
then the previous identities imply the equations (Bel, Lichnerowicz) 
\begin{equation}
\nabla _{{\alpha }}R^{{\alpha }}\mathstrut _{{\beta },{\lambda }\mu }=\nabla
_{{\lambda }}\rho _{\mu {\beta }}-\nabla _{\mu }\rho _{{\lambda }{\beta }}.
\label{11.6}
\end{equation}
Equations (\ref{11.2}) and (\ref{11.6}) are analogous to the Maxwell
equations for the electromagnetic 2-form $F$: 
\begin{equation}
dF=0,\text{ \ \ }\delta F=J,
\end{equation}
where $J$ is the electric current.

\begin{theorem}
The Riemann tensor of an Einsteinian spacetime of arbitrary dimension
satisfies a quasi-diagonal, semilinear system of wave equations.
\end{theorem}

\begin{proof}
One deduces from (\ref{11.2}) and the Ricci identity, an identity of the
form: 
\begin{equation}
\nabla ^{\alpha }\nabla _{{\alpha }}R_{{\beta }{\gamma },{\lambda }\mu
}+\nabla _{{\gamma }}\nabla ^{\alpha }R_{{\alpha }{\beta },{\lambda }\mu
}+\nabla _{{\beta }}\nabla ^{\alpha }R_{{\gamma }{\alpha },{\lambda }\mu
}+S_{\beta \gamma ,\lambda \mu }\equiv 0,
\end{equation}
where $S_{\beta \gamma ,\lambda \mu }$ is an homogeneous quadratic form in
the Riemann tensor, 
\begin{equation}
S_{\beta \gamma ,\lambda \mu }\equiv \{R_{\gamma }{}^{\rho }R_{\rho \beta
,\lambda \mu }+R^{\alpha }{}_{\gamma ,\beta }{}^{\rho }R_{\alpha \rho
,\lambda \mu }+[(R^{\alpha }{}_{\gamma },_{\lambda }{}^{\rho }R_{\alpha
\beta ,\rho \mu })-(\lambda \rightarrow \mu )]\}-\{\beta \rightarrow \gamma
\}.
\end{equation}
Using equations (\ref{11.6}), when the Ricci tensor satisfies the Einstein
equations, gives equations of the form, 
\begin{equation}
\nabla ^{\alpha }\nabla _{{\alpha }}R_{{\beta }{\gamma },{\lambda }\mu
}+S_{\beta {\gamma ;\lambda \mu }}=J_{{\beta \gamma ,\lambda \mu }},
\label{11.10}
\end{equation}
with $J_{\beta \gamma ,\lambda \mu }$ depending on the sources $\rho
_{\alpha \beta \text{ }}$ and being zero in vacuum: 
\begin{equation}
J_{\beta \gamma ,\lambda \mu }\equiv \nabla _{\gamma }(\nabla _{\mu }\rho _{{%
\lambda }{\beta }}-\nabla _{{\lambda }}\rho _{\mu {\beta }})-(\beta
\rightarrow \gamma ),  \label{11.11}
\end{equation}
and this completes the proof.
\end{proof}

\subsection{Case n=3, FOS system}

In a coframe $\theta ^{0},\theta ^{i}$ where $g_{0i}=0$, equations (\ref
{11.2}) with $\{{\alpha }{\beta }{\gamma }\}=\{ijk\}$ and equations (\ref
{11.6}) with ${\ \beta }=0$ do not contain derivatives $\partial _{0}$ of
the Riemann tensor. We call them `Bianchi constraints'. The remaining
equations, called from here on `Bianchi equations,' read as follows: 
\begin{equation}
\nabla _{0}R_{hk,{\lambda }\mu }+\nabla _{k}R_{0h,{\lambda }\mu }+\nabla
_{h}R_{k0,{\lambda }\mu }=0,
\end{equation}
\begin{equation}
\nabla _{0}R^{0}\mathstrut _{i,{\lambda }\mu }+\nabla _{h}R^{h}\mathstrut
_{i,{\lambda }\mu }=\nabla _{{\lambda }}\rho _{\mu i}-\nabla _{\mu }\rho _{{%
\lambda }i}\equiv J_{{\lambda }\mu i},
\end{equation}
where the pair $({\lambda }\mu )$ is either $(0j)$ or $(jl),$ with $j<l$.
There are 3 of one or the other of these pairs if the space dimension $n$ is
equal to 3.

Equations (\ref{11.12}) and (\ref{11.13}) are, for each given pair $(0j)$, a
first order system for the components $R_{hk,0j}$ and $R_{0h,0j}.$ If we
choose at a point of the spacetime an orthonormal frame the principal
operator is diagonal by blocks, each block corresponding to a choice of a
pair ($\lambda \mu ,\lambda <\mu ),$ is a symmetric 6 by 6 matrix which
reads:

\begin{center}
$\left( \bigskip 
\begin{tabular}{cccccc}
$\partial_{0}$ & 0 & 0 & $\partial_{2}$ & -$\partial_{1}$ & 0 \\ 
0 & $\partial_{0}$ & 0 & 0 & $\partial_{3}$ & -$\partial_{2}$ \\ 
0 & 0 & $\partial_{0}$ & -$\partial_{3}$ & 0 & $\partial_{1}$ \\ 
$\partial_{2}$ & 0 & -$\partial_{3}$ & $\partial_{0}$ & 0 & 0 \\ 
-$\partial_{1}$ & $\partial_{3}$ & 0 & 0 & $\partial_{0}$ & 0 \\ 
0 & -$\partial_{2}$ & $\partial_{1}$ & 0 & 0 & $\partial_{0}$%
\end{tabular}
\right) $.
\end{center}

We have proved:

\begin{theorem}
The Bianchi evolution equations are a FOS (first order symmetrizable) system.
\end{theorem}

The Bianchi equations depend on the choice of frame, as does their
hyperbolicity.

\subsection{Cauchy adapted frame}

The numerical valued matrix $\mathcal{M}^{t}$ of coefficients of the
operator $\partial/\partial t$ corresponding to the Bianchi equations
relative to the Cauchy adapted frame is proportional to the unit matrix,
with coefficient $N^{-2},$ hence is positive definite and the following
theorem holds.

\begin{theorem}
The Bianchi equations associated to a Cauchy adapted frame are a FOSH
system, with space sections $\mathcal{M}_{t}$.
\end{theorem}

We will give an explicit expression of the full system after introducing two
`electric' and two `magnetic' space tensors associated with the double
2-form $R.$ They are the gravitational analogs of the electric and magnetic
vectors associated with the electromagnetic 2-form $F.$ That is, we define
the `electric' tensors by, 
\begin{equation}
E_{ij}\equiv R^{0}\mathstrut_{i,0j},  \label{electric}
\end{equation}
\begin{equation}
D_{ij}\equiv{\frac{1}{4}}\eta_{ihk}\eta_{jlm}R^{hk,lm},
\end{equation}
while the `magnetic' tensors are given by, 
\begin{eqnarray}
H_{ij} & \equiv&{\frac{1}{2}}N^{-1}\eta_{ihk}R^{hk}\mathstrut_{,0j}, \\
B_{ji} & \equiv&{\frac{1}{2}}N^{-1}\eta_{ihk}R_{0j,}\mathstrut^{hk}.
\label{magnetic}
\end{eqnarray}
In these formulae, $\eta_{ijk}$ is the volume form of the space metric $\bar{
g}\equiv g_{ij}dx^{i}dx^{j}$.

\begin{lemma}
\begin{enumerate}
\item  The electric and magnetic tensors are always such that 
\begin{equation}
E_{ij}=E_{ji},\text{ \ \ }D_{ij}=D_{ji},\text{ \ \ }H_{ij}=B_{ji}
\end{equation}

\item  If the Ricci tensor satisfies the vacuum Einstein equations with
cosmological constant 
\begin{equation}
R_{\alpha \beta }=\Lambda g_{\alpha \beta }
\end{equation}
then the following additional properties hold 
\begin{equation}
H_{ij}=H_{ji}=B_{ij}=B_{ji},\text{ \ \ }E_{ij}=D_{ij}  \label{11.18}
\end{equation}
\end{enumerate}
\end{lemma}

\begin{proof}
(1) The Riemann tensor is a symmetric double 2-form, the electric and
magnetic 2-tensors associated to it by the relations possess obviously the
given symmetries.

(2) The Lanczos identity for a symmetric double two-form, with a tilde
representing the spacetime double dual, gives 
\begin{equation}
\tilde{R}_{{\alpha }{\beta },{\lambda }\mu }+R_{{\alpha }{\beta },{\lambda }%
\mu }\equiv C_{{\alpha }{\lambda }}g_{{\beta }\mu }-C_{{\alpha }\mu }g_{{%
\beta }{\lambda }}+C_{{\beta }\mu }g_{{\alpha }{\lambda }}-C_{{\beta }{%
\lambda }}g_{{\alpha }\mu },
\end{equation}
with 
\begin{equation}
C_{{\alpha }{\beta }}\equiv R_{{\alpha }{\beta }}-{\frac{1}{4}}Rg_{{\alpha }{%
\beta }}.
\end{equation}
It implies that $\tilde{R}_{{\alpha }{\beta },{\lambda }\mu }+R_{{\alpha }{%
\beta },{\lambda }\mu }=0$ if $C_{\alpha \beta }=0$, in particular for an
Einsteinian vacuum spacetime with possibly a cosmological constant. The
relations (\ref{11.18}) can then be proved by a straightforward calculation
that employs the relation $\eta _{0ijk}=N\eta _{ijk}$ between the spacetime
and space volume forms.
\end{proof}

In order to extend the treatment to the non-vacuum case and to avoid
introducing unphysical characteristics in the solution of the Bianchi
equations, we will keep as independent unknowns the four tensors $\mathbf{E}$
, $\mathbf{D}$, $\mathbf{B}$, and $\mathbf{H}$, which will not be regarded
necessarily as symmetric. The symmetries will be imposed eventually on the
initial data and shown to be conserved by evolution.

We now express the Bianchi equations in terms of the time-dependent space
tensors $\mathbf{E}$, $\mathbf{H}$, $\mathbf{D}$, and $\mathbf{B}$. We use
the following relations, found by inverting the definitions (\ref{electric}%
)-(\ref{magnetic}), 
\begin{align}
R_{0i,0j}& =-N^{2}E_{ij},\text{ \ \ }R_{hk,0j}=N\eta ^{i}\mathstrut
_{hk}H_{ij}, \\
R_{hk,lm}& =\eta ^{i}\mathstrut _{hk}\eta ^{j}\mathstrut _{lm}D_{ij},\text{
\ \ }R_{0j,lm}=N\eta ^{i}\mathstrut _{lm}B_{ji}.
\end{align}
We will express spacetime covariant derivatives of the Riemann tensor in
terms of space covariant derivatives ${\bar{\nabla}}$ and time derivatives, $%
{\ \hat{\partial}_{0}}$, of $\mathbf{E}$, $\mathbf{H}$, $\mathbf{D}$, $%
\mathbf{B} $ by using the connection coefficients in (3+1)-form as given in
Section \ref{3}.

The first Bianchi equation with $[{\lambda }\mu ]=[0j]$ has the form, 
\begin{equation}
\nabla _{0}R_{hk,0j}+\nabla _{k}R_{0h,0j}-\nabla _{h}R_{0k,0j}=0.
\end{equation}
A calculation incorporating previous definitions, then grouping derivatives
using ${\hat{\partial}_{0}}$ and ${\bar{\nabla}}_{i}$, gives to the first
pair of Bianchi equations, with $[{\lambda }\mu ]=[0j],$ the following
forms: 
\begin{equation}
{\hat{\partial}_{0}}E_{ij}-N\eta ^{hl}\mathstrut _{i}{\bar{\nabla}}%
_{h}H_{lj}+(L_{2})_{ij}=J_{0ji},  \label{11.24}
\end{equation}
where $\mathbf{J}$ is zero in vacuum and 
\begin{equation}
{\hat{\partial}_{0}}(\eta ^{i}\mathstrut _{hk}H_{ij})+2N{\bar{\nabla}}%
_{[h}E_{k]j}+(L_{1})_{hk,j}=0,  \label{11.25}
\end{equation}
with, 
\begin{align}
(L_{2})_{ij}& \equiv -N({\mathrm{tr}}K)E_{ij}+NK^{k}\mathstrut
_{j}E_{ik}+2NK_{i}\mathstrut ^{k}E_{kj} \\
& -({\bar{\nabla}}_{h}N)\eta ^{hl}\mathstrut _{i}H_{lj}+NK^{k}\mathstrut
_{h}\eta ^{lh}\mathstrut _{i}\eta ^{m}\mathstrut _{kj}D_{lm}+({\bar{\nabla}}%
^{k}N)\eta ^{l}\mathstrut _{kj}B_{il},
\end{align}
\begin{align}
(L_{1})_{hk,j}& \equiv NK^{l}\mathstrut _{j}\eta ^{i}\mathstrut
_{hk}H_{il}+2({\bar{\nabla}}_{[h}N)E_{k]j}+2N\eta ^{i}\mathstrut
_{lj}K^{l}\mathstrut _{\lbrack k}B_{h]i} \\
& -({\bar{\nabla}}^{l}N)\eta ^{i}\mathstrut _{hk}\eta ^{m}\mathstrut
_{lj}D_{im}.
\end{align}
We see that the non-principal terms $L_{1}$ and $L_{2}$ are linear in $%
\mathbf{E}$, $\mathbf{D}$, $\mathbf{B}$, and $\mathbf{H}$, with coefficients
linear in the geometrical elements $\mathbf{K}$ and ${\mathbf{\bar{\nabla}}}%
N $. The characteristic matrix of the principal terms is symmetrizable. The
unknowns $E_{i(j)}$ and $H_{i(j)}$, with fixed $j$ and $i=1,2,3$ appear only
in the equations with given $j$. The other unknowns appear in non-principal
terms. The characteristic matrix is composed of three blocks around the
diagonal, each corresponding to one given $j$.

The $j^{\mathrm{th}}$ block of the characteristic matrix in an orthonormal
frame for the space metric ${\bar{\mathbf{g}}}$, with unknowns listed
horizontally and equations listed vertically ($j$ is suppressed), is given
by, 
\begin{equation}
\left( 
\begin{array}{cccccc}
\xi_{0} & 0 & 0 & 0 & N\xi_{3} & -N\xi_{2} \\ 
0 & \xi_{0} & 0 & -N\xi_{3} & 0 & N\xi_{1} \\ 
0 & 0 & \xi_{0} & N\xi_{2} & -N\xi_{1} & 0 \\ 
0 & -N\xi_{3} & N\xi_{2} & \xi_{0} & 0 & 0 \\ 
N\xi_{3} & 0 & -N\xi_{1} & 0 & \xi_{0} & 0 \\ 
-N\xi_{2} & N\xi_{1} & 0 & 0 & 0 & \xi_{0}
\end{array}
\right) .
\end{equation}
This matrix is symmetric and its determinant is the characteristic
polynomial of the $\mathbf{E}$, $\mathbf{H}$ system. It is given by, 
\begin{equation}
-N^{6}(\xi_{0}\xi^{0})(\xi_{{\alpha}}\xi^{{\alpha}})^{2}.
\end{equation}
The characteristic matrix is symmetric in an orthonormal space frame and the
coefficient matrix $\mathcal{M}^{t}$ is positive definite (it is the unit
matrix). Therefore, the first order system is symmetrizable hyperbolic with
respect to the space sections $\mathcal{M}_{t}$. We do not have to compute
the symmetrized form explicitly because one can obtain energy estimates
directly by using the contravariant associates $E^{ij},\, H^{ij},\ldots$ of
the unknowns.

The second pair of Bianchi equations, for $D_{ij}$ and $B_{ij},$ obtained
for $[{\lambda}\mu]=[lm]$ is analogous. The characteristic matrix for the $%
[lm]$ equations, with unknowns $D_{ij}$ and $B_{ij}$, $j$ fixed, with an
orthonormal space frame, is the same as the matrix found above.

If the spacetime metric $\mathbf{g}$ is considered as given, as well as the
sources, the Bianchi equations form a linear symmetric hyperbolic system
with domain of dependence determined by the light cone of $\mathbf{g}$. The
coefficients of the terms of order zero are ${\mathbf{\bar{\nabla}}}N$ or $N%
\mathbf{K}$. The system is homogeneous in vacuum (zero sources).

\subsection{FOSH system for $\mathbf{u}\equiv(\mathbf{E},\mathbf{H},\mathbf{D%
},\mathbf{B},{\bar{\mathbf{g}}},\mathbf{K},\bar{\Gamma})$}

The Bianchi equations depend on the metric. Our problem is to find a system
for determining the metric from the Riemann tensor (through eventually other
auxiliary unknowns), which together with the Bianchi equations, constitute a
well-posed system. It is possible to construct a FOSH system linking the
metric and the connection to our Bianchi field (Anderson, C.B and York
1997), if we again densitize the lapse, i.e., set $N={\alpha }(\det {\bar{%
\mathbf{g}} })^{1/2}$. This system is inspired by an analogous one
constructed in conjunction with the Weyl tensor by H. Friedrich 1996.

\subsection{Elliptic - hyperbolic system}

Instead of determining the metric from the curvature through hyperbolic
equations, one can try to do so by elliptic equations on space slices.
Well-posed problems for such equations are essentially global ones and
depend on the global geometric properties of the space manifolds.

\subsubsection{Determination of $K$}

We deduce from the identities (\ref{3.17}), which read, 
\begin{equation}
N^{-1}R_{0i,jk}\equiv {\bar{\nabla}}_{j}K_{ki}-{\bar{\nabla}}_{k}K_{ji},
\end{equation}
and the Ricci identity, that 
\begin{equation}
\bar{\nabla}^{j}(N^{-1}R_{0i,jk})\equiv \bar{\nabla}^{j}{\bar{\nabla}}
_{j}K_{ki}-\bar{\nabla}_{k}{\bar{\nabla}}^{j}K_{ji}-\bar{R} _{\dots\,
kjh}^{j}K_{hi}-\bar{R}_{\dots\, kih}^{j}K_{ji}.
\end{equation}
As a gauge condition we now suppose that $H\equiv K_{h}^{h}$ is a given
function $h$ on the spacetime, $N$ then satisfies on each slice the equation
deduced from the Einstein equation $R_{00}=0$: 
\begin{equation}
\bar{\Delta}N-K^{ij}K_{ij}N=-\partial _{0}h
\end{equation}
The use of the momentum constraint on each slice gives for $K$ (when $N$ and
the Riemann tensor of spacetime are known) a quasi-diagonal semilinear
system, elliptic if $\bar{g}$ is properly Riemannian, namely, 
\begin{equation}
\bar{\nabla}^{j}{\bar{\nabla}}_{j}K_{ki}-\bar{R}_{\dots\,kjh}^{j}K_{hi}-\bar{%
R} _{\dots\, kih}^{j}K_{ji}=\bar{\nabla}_{k}\partial _{i}h+\bar{\nabla}
^{j}(N^{-1}R_{0i,jk}),
\end{equation}
where (cf. (\ref{3.16})), 
\begin{equation}
\bar{R}_{ij,kl}=R_{ij,kl}-K_{ik}K_{lj}+K_{il}K_{kj}.
\end{equation}
The global solvability of the equation for $K$ can be proved under some
conditions, for instance in a neighborhood of Euclidean space, perhaps also
for a Robertson-Walker spacetime with compact space of negative curvature.

\subsubsection{Determination of $\bar{g}$}

The equation $\hat{\partial}_{0}g_{ij}=-2NK_{ij}$ determines as before $\bar{%
g}$ when $N,\beta $ and $K$ are known. However it does not improve the
regularity on $\mathcal{M}_{t}$ of $\bar{g}$ over the regularity of $K.$

A better result can be sought through the identity which gives the Ricci
tensor of $\bar{g}$ in terms of the Riemann tensor of spacetime and $K$ by, 
\begin{equation}
\bar{R}_{ij}\equiv \bar{g}^{hk}R_{ih,jk}+HK_{ij}-K_{i}^{h}K_{jh}.
\end{equation}
Various methods have been devised to determine a Riemannian metric from its
Ricci tensor by elliptic equations (see in particular Andersson and
Moncrief, to appear, for the case of compact manifolds with negative
curvature).

\section{Bel-Robinson energy}

The new formulation brings nothing new for the local existence and
uniqueness theorem. It is useful in obtaining geometrical energy estimates
(Bel-Robinson energy) leading possibly to global existence theorems. Such
estimates have been used by Christodoulou and Klainerman 1989 in the case of
asymptotically Euclidean manifolds. They are used by Andersson and Moncrief
(to appear) for compact manifolds with negative curvature.

\subsection{Bel-Robinson energy in a strip}

Multiply (\ref{11.25}) by ${\frac{1}{2}}\eta _{l}\mathstrut ^{hk}H^{lj}$,
and recall that $\eta _{l}\mathstrut ^{hk}\eta ^{i}\mathstrut _{hk}=2\delta
^{i}\mathstrut _{l}$, $\eta _{lrk}\eta ^{ihk}=\delta ^{i}\mathstrut
_{l}\delta ^{h}\mathstrut _{r}-\delta ^{i}\mathstrut _{r}\delta
^{h}\mathstrut _{l}$, and ${\hat{\partial}_{0}}g^{ij}=2NK^{ij}$. Then we
find that, 
\begin{equation}
{\frac{1}{2}}\eta _{l}\mathstrut ^{hk}H^{lj}{\hat{\partial}_{0}}(\eta
^{i}\mathstrut _{hk}H_{ij})={\frac{1}{2}}{\hat{\partial}_{0}}%
(H_{ij}H^{ij})-M_{1},
\end{equation}
\begin{align}
M_{1}& \equiv {\frac{1}{4}}\eta ^{l}\mathstrut _{rs}H_{lm}\eta
^{i}\mathstrut _{hk}H_{ij}{\hat{\partial}_{0}}(g^{hr}g^{ks}g^{jm}) \\
& =N(K_{h}^{h}H^{ij}-K^{i}\mathstrut _{l}H^{lj}+K_{l}\mathstrut
^{j}H^{il})H_{ij}.  \notag
\end{align}
Likewise, multiply (\ref{11.24}) by $E^{ij}$ to obtain, 
\begin{equation}
E^{ij}{\hat{\partial}_{0}}E_{ij}={\frac{1}{2}}{\hat{\partial}_{0}}%
(E_{ij}E^{ij})-M_{2},
\end{equation}
\begin{equation}
M_{2}\equiv N(K^{i}\mathstrut _{l}E^{lj}+K_{l}\mathstrut ^{j}E^{il})E_{ij}.
\end{equation}
Multiplication by appropriate factors (cf. Anderson, C.B and York 1997) of
the second pair of Bianchi equations leads to analogous results. The sum of
the expressions so obtained from the four Bianchi equations gives an
expression where the spatial derivatives add to form an exact spatial
divergence, just as for all symmetric systems. Indeed, we obtain, 
\begin{eqnarray}
&\frac{1}{2}&{\hat{\partial}_{0}}\left( |\mathbf{E}|^{2}+|\mathbf{H}|^{2}+|%
\mathbf{D}|^{2}+|\mathbf{B}|^{2}\right) +{\bar{\nabla}}_{h}(NE^{ij}\eta
^{lh}\mathstrut _{i}H_{lj})  \notag \\
&-&\bar{\nabla}_{h}(NB^{ij}\eta ^{lh}\mathstrut _{i}D_{lj})=Q(\mathbf{E},%
\mathbf{H},\mathbf{D},\mathbf{B})+\mathcal{S},  \label{12.5}
\end{eqnarray}
where we have denoted by $|\cdot |$ the pointwise ${\bar{\mathbf{g}}}$ norm
of a space tensor, and where $Q$ is a quadratic form with coefficients ${%
\mathbf{\bar{\nabla}}}N$ and $\mathbf{K.}$ The source term $\mathcal{S}$,
zero in vacuum, is, 
\begin{equation}
\mathcal{S}\equiv J_{0ij}E^{ij}-{\frac{1}{2}}NJ_{lmi}\eta _{h}\mathstrut
^{lm}B^{ih}.
\end{equation}
We define the \textit{Bel-Robinson energy} at time $t$ of the field $(%
\mathbf{E},\mathbf{H},\mathbf{D},\mathbf{B})$, called a `Bianchi field' when
it satisfies the Bianchi equations, to be the integral, 
\begin{equation}
\mathcal{B}(t)\equiv {\frac{1}{2}}\int_{\mathcal{M}_{t}}(|\mathbf{E}|^{2}+|%
\mathbf{H}|^{2}+|\mathbf{D}|^{2}+|\mathbf{B}|^{2})\mu _{{\bar{\mathbf{g}}}%
_{t}}.
\end{equation}
We will prove the following.

\begin{theorem}
Suppose that $\mathbf{g}$ is $\mathcal{C}^{1}$ on $\mathcal{M}\times \lbrack
0,T]$ and that the ${\bar{\mathbf{g}}}_{t}$ norms of ${\mathbf{\bar{\nabla}}}%
N$ and $\mathbf{K}$ are uniformly bounded on $\mathcal{M}_{t}$, $t\in
\lbrack 0,T]$. Denote by $\pi (t)$ the supremum 
\begin{equation}
\pi (t)={\mathrm{Sup}}_{\mathcal{M}_{t}}(|{\mathbf{\bar{\nabla}}}N|+|\mathbf{%
K}|).
\end{equation}
Suppose the matter source $\mathbf{J}\in L^{1}([0,T],L^{2}(\mathcal{M}_{t}))$%
, then the Bel energy of a $\mathcal{C}^{1}$ Bianchi field with compact
support in space satisfies for $0\leq t\leq T$ the following inequality, 
\begin{equation}
\mathcal{B}(t)^{1/2}\leq (\mathcal{B}(0)^{1/2}+\frac{C}{2}\int_{0}^{t}\Vert
J\Vert _{L^{2}(\mathcal{M}_{\tau })}d\tau )\exp (C\int_{0}^{t}\pi (\tau
)d\tau ),
\end{equation}
where $C$ is a given positive number.
\end{theorem}

\begin{proof}
We integrate the identity (\ref{12.5}) above on the strip $\mathcal{M}\times
\lbrack 0,t]$ with respect to the volume element $\mu _{{\bar{\mathbf{g}}}%
_{\tau }}d\tau $. If the Bianchi field has support compact in space the
integral of the space divergence term vanishes. The integration of ${%
\partial _{0}}f$ on a strip of spacetime with respect to the volume form $%
d\tau \,\mu _{{\bar{\mathbf{g}}}_{\tau }}$ goes as follows, for an arbitrary
function $f,$ 
\begin{equation}
\int_{0}^{t}\int_{\mathcal{M}_{\tau }}\partial {_{0}}f\,\mu _{{\bar{\mathbf{g%
}}}_{\tau }}d\tau \equiv \int_{0}^{t}\int_{\mathcal{M}_{\tau }}({\partial }%
_{t}-\beta ^{i}{\partial }_{i})f\,\mu _{{\bar{\mathbf{g}}}_{\tau }}\,d\tau .
\notag
\end{equation}
It holds that, 
\begin{equation}
\int_{\mathcal{M}_{t}}\partial {_{0}}f\,\mu _{{\bar{\mathbf{g}}}_{t}}=\int_{%
\mathcal{M}_{t}}\{{\partial }_{t}(f\mu _{\bar{g}_{t}})\,-f{\partial }%
_{t}\,\mu _{{\bar{\mathbf{g}}}_{t}}-{\bar{\nabla}}_{i}(\beta ^{i}f)\,-f{\bar{%
\nabla}}_{i}\beta ^{i}\}\,\mu _{{\bar{\mathbf{g}}}_{t}}.
\end{equation}
Using the expression for the derivative of a determinant and the relation
between $\bar{g}$ and $K,$ we find that, 
\begin{equation}
\partial _{t}\mu _{\bar{g}_{t}}=(-N{\mathrm{tr}}\mathbf{K}+\bar{\nabla}%
_{i}\beta ^{i})\mu _{\bar{g}_{t}}
\end{equation}
and therefore if $f$ has compact support in space, 
\begin{equation}
\int_{\mathcal{M}_{t}}\partial {_{0}}f\,\mu _{{\bar{\mathbf{g}}}_{t}}={%
\partial }_{t}\int_{\mathcal{M}_{t}}f\mu _{\bar{g}_{t}}+\int_{\mathcal{M}%
_{t}}fN{\mathrm{tr}}\mathbf{K}\,\mu _{{\bar{\mathbf{g}}}_{t}}.
\end{equation}
The integration on a strip leads therefore to the equality, 
\begin{equation}
\mathcal{B}(t)=\mathcal{B}(0)+\int_{0}^{t}\int_{\mathcal{M}_{\tau }}(\tilde{Q%
}+\mathcal{S})\,\mu _{{\bar{\mathbf{g}}}_{\tau }}\,d\tau ,
\end{equation}
with\footnote{{\footnotesize \ One can take advantage of the decomposition
of \~{Q} to obtain better estimates in the case TrK}$\leq 0.$}, 
\begin{equation*}
\tilde{Q}=Q+{\frac{1}{2}}N({\mathrm{tr}}\mathbf{K})(|\mathbf{E}|^{2}+|%
\mathbf{H}|^{2}+|\mathbf{D}|^{2}+|\mathbf{B}|^{2}).
\end{equation*}
We deduce from this equality, and the expression indicated in (\ref{12.5})
for $Q$, the following inequality, with $C$ some number, 
\begin{equation}
\mathcal{B}(t)\leq \mathcal{B}(0)+C\{\int_{0}^{t}{\mathrm{Sup}}_{M_{\tau }}(|%
{\mathbf{\bar{\nabla}}}N|+|\mathbf{K}|)\mathcal{B}(\tau )\,+\Vert J\Vert
_{L^{2}(M_{\tau })}\mathcal{B}(t)^{1/2}\}\,d\tau .
\end{equation}
This inequality and the resolution of the corresponding equality imply the
result.
\end{proof}

\begin{remark}
The quantities $-{\bar{\nabla}}_{k}N$ and $2K_{ij}$ are respectively the $%
(0k)$ and $(ij)$ components of the Lie derivative $\mathcal{L}_{n}\mathbf{g}$
of the spacetime metric $\mathbf{g}$ with respect to the unit normal \textbf{%
n} to $\mathcal{M}_{t}$ (its $(00)$ component is identically zero). The
Bel-Robinson energy is therefore conserved if this Lie derivative is zero.
\end{remark}

The estimate of the Bel-Robinson energy is only an intermediate step in
global existence proofs since it depends on the metric which itself depends
on curvature.

\subsection{Local energy estimate}

We take as a domain $\Omega $ of spacetime the closure of a connected open
set whose boundary ${\partial }\Omega $ consists of three parts: A domain ${%
\ \omega }_{t}$ of $\mathcal{M}_{t}$, a domain ${\omega }_{0}$ of $\mathcal{M%
}_{0}$, and a lateral boundary $\mathcal{L}$. We assume $\mathcal{L}$ is
spacelike or null and `ingoing', that is timelike lines entering $\Omega $
at a point of $\mathcal{L}$ are past-directed. We also assume that the
boundary ${\partial }\Omega $ is regular in the sense of Stokes formula. We
use the identity previously found and integrate this identity on $\Omega $
with respect to the volume form $\mu _{{\bar{\mathbf{g} }}_{\tau }}\,d\tau $%
. It can be proved that the integral on $\mathcal{L}$ resulting from the
application of Stokes formula is nonnegative. The Bel-Robinson energy on ${%
\omega }_{t}$ satisfies therefore the same type of inequality as found
before on $\mathcal{M}_{t}.$ In particular, we have $\mathcal{B}({\omega }%
_{\tau })=0$, if $\mathcal{B}({\omega }_{0})=0$ and $\mathbf{J}=0$ (vacuum
case). Then $\mathbf{E}=\mathbf{H}=\mathbf{D}=\mathbf{B}=0$ in $\Omega $ if
they vanish on the intersection of $\mathcal{M}_{0}$ with the past of $%
\Omega $ (result found by York 1987). Note that such a result is not
sufficient to prove the propagation of gravitation with the speed of light
because it treats only curvature tensors that are zero in some domain, not
the difference of nonzero curvature tensors. The Bianchi equations are not
by themselves sufficient to estimate such differences because their
coefficients depend on the metric, which itself depends on the curvature.

\section{(n+1)-splitting in a time-adapted frame}

\subsection{Metric and coframe}

We choose the time axis to be \emph{tangent to the time lines}, i.e., the
cobasis $\theta$ is such that $\theta^{i}$ does not contain $dx^{0}.$ We
set, 
\begin{equation}
\theta^{i}=a_{j}^{i}dx^{j},\text{ \ \ \ \ }\theta^{0}=Udx^{0}+b_{i}dx^{i}.
\end{equation}
We will call such a frame a \textsc{CF}- (Cattaneo-Ferrarese) frame. The
Pfaff derivatives $\partial_{\alpha}$ in the \textsc{CF}-frame are linked to
the partial derivatives $\partial/\partial x^{\alpha}$ by the relations, 
\begin{equation*}
\partial_{0}=U^{-1}\frac{\partial}{\partial x^{0}},\text{ \ \ }\partial
_{i}=A_{i}^{j}\left[\frac{\partial}{\partial x^{j}}-U^{-1}b_{j}\frac{%
\partial }{ \partial x^{0}}\right].\,
\end{equation*}
with $A_{i}^{j}$ the matrix inverse of $a_{i}^{j}.$ The structure
coefficients of the coframe are found to be, 
\begin{equation*}
c_{0i}^{0}=N^{-1}\partial_{i}N-A_{i}^{j}\partial_{0}b_{j}=-c_{i0}^{0},
\end{equation*}
and with $f_{[ij]}\equiv f_{ij}-f_{ji},$ 
\begin{equation*}
c_{ij}^{0}=UA_{[i}^{h}\partial_{j]}(N^{-1}b_{h}),\text{ }%
c_{k0}^{i}=A_{k}^{j}\partial_{0}a_{j}^{i},\text{ }c_{hk}^{i}=A_{[h}^{j}
\partial_{k]}a_{j}^{i}.
\end{equation*}

\begin{remark}
If the time lines are not hypersurface orthogonal (i.e., if $b_{i}\not =0)$,
the coefficients $c_{ij}^{h}$ are different from the structure coefficients
of the space frame $\theta^{i}.$
\end{remark}

Choosing the frame to be orthonormal the metric reads, 
\begin{equation}
g=-(\theta ^{0})^{2}+\sum_{i=1}^{3}(\theta ^{i})^{2}.
\end{equation}

\subsection{Splitting of connection}

We deduce from the general formulas, 
\begin{equation*}
\omega_{00}^{0}=\omega_{i0}^{0}=0,
\end{equation*}
\begin{equation*}
Y_{i}\equiv\omega_{00,i}=\omega_{0i}^{0}=%
\omega_{00}^{i}=-c_{0i,0}=c_{0i}^{0}=U^{-1}\partial_{i}U-A_{i}^{j}
\partial_{0}b_{j},
\end{equation*}
and we know that $\omega_{0i}^{j}=\omega_{oi,j}$ is antisymmetric in $i$ and 
$j.$ We set, 
\begin{equation*}
\omega_{0i,j}\equiv f_{ij}=\frac{1}{2}\{A_{j}^{h}%
\partial_{0}a_{h}^{i}-A_{i}^{h}\partial_{0}a_{i}^{j}+A_{[i}^{h}
\partial_{j]}b_{h}\}.
\end{equation*}
Let $e_{\alpha}\equiv\partial_{\alpha}$ be the frame dual to $%
\theta^{\alpha} $, i.e., such that the vector $e_{\alpha}$ has components $%
\delta_{(\alpha)}^{\lambda}.$ Then, 
\begin{equation*}
\nabla_{\beta}e_{\alpha}^{\lambda}=\omega_{\beta\alpha}^{\lambda},
\end{equation*}
in particular, $\omega_{0i,j}$ is the projection on $e_{(j)}$ of the
derivative of $e_{(i)}$ in the direction of $e_{(0)}.$ We have $f_{ij}=0$ if
the frame is Fermi-transported along the time line. We will make this
hypothesis to simplify the formulas.

The connection coefficient $\omega_{i0,j}$ is the sum of a term symmetric in 
$i$ and $j$ and an antisymmetric one and therefore we have, 
\begin{equation*}
X_{ij}\equiv\omega_{i0,j}=\omega_{i0}^{j}=\frac{1}{2}\{A_{j}^{h}\partial
_{0}a_{h}^{i}+A_{i}^{h}\partial_{0}a_{i}^{j}+A_{[i}^{h}\partial_{j]}b_{h}\}.
\end{equation*}
The antisymmetric term vanishes if the time lines are hypersurface
orthogonal ($b_{i}=0).$

The coefficients $\omega_{ij}^{h}$ are also linear expressions in terms of
the first derivatives of the frame coefficients, they are identical to the
connection constructed with the $a_{i}^{j}$ at fixed $t$ only if $b_{i}=0.$

\subsection{Splitting of curvature}

Using the general formulas we find in the chosen frame, 
\begin{equation}
R_{0h\dots\, j}^{\dots\, i}\equiv \partial _{0}\omega
_{hj}^{i}+X_{h}^{\dots\, k}\omega
_{kj}^{i}+Y^{i}X_{hj}^{\dots}-Y_{j}X_{h}{}^{i}
\end{equation}
we denote by $\tilde{\nabla}$ is the pseudo-covariant derivative constructed
with $\partial _{i}$ and $\omega _{ij}^{h}$ (Cataneo-Ferrarese transversal
derivative). We have, 
\begin{equation}
R_{h0..0}^{\dots\, i}\equiv \tilde{\nabla}_{h}Y^{i}-\partial
_{0}X_{h}^{\dots\, i}-X_{h}^{\dots\, j}X_{j}^{\dots\, i}
\end{equation}
\begin{equation}
R_{hk\dots\, j}^{\dots\, i}\equiv\tilde{R} _{hk\dots\, j}^{\dots\,
^{i}}+X_{\dots\, k}^{\dots\, i}X_{jh}-X_{jk}^{\dots\,}X_{\dots\, h}^{i},
\end{equation}
where $\tilde{R}_{hk\dots\, j}^{\dots\, ^{i}}$ denotes the expression
formally constructed as a Riemann tensor with the coefficients $%
\omega_{ij}^{h}$ and, 
\begin{equation}
R_{kh\dots\, j}^{\dots\, 0}\equiv\tilde{\nabla}_{k}X_{hj}-\tilde{\nabla}
_{h}X_{kj}-Y_{j}(X_{kh}-X_{hk}).
\end{equation}

\begin{remark}
The symmetry $R_{kh,0j}=R_{0j,kh}$ results from the expression of the
connection in terms of frame coefficients.
\end{remark}

We deduce from the splitting of the Riemann tensor the following identities: 
\begin{equation}
R_{00}\equiv\tilde{\nabla}_{i}Y^{i}- \partial_{0}X_{i}^{i}-X_{h}^{\dots\,
j}X_{j}^{\dots\, h},
\end{equation}
\begin{equation*}
R_{h0}\equiv\tilde{\nabla}_{j}X_{h}^{\dots\, j}-\tilde{\nabla}
_{h}X_{j}^{j}-Y^{j}(X_{jh}-X_{hj}).
\end{equation*}

\subsection{Bianchi equations (case n=3)}

\subsubsection{Bianchi quasi constraints}

The Bianchi identities and their contraction contain, as in a Cauchy adapted
frame, equations which do not contain the derivative $\partial_{0}$ of the
Riemann tensor, namely, 
\begin{equation}
\nabla_{i}R_{jh,\lambda\mu}+\nabla_{h}R_{ij,\lambda\mu}+\nabla_{j}R_{hi,
\lambda\mu}\equiv0,
\end{equation}
\begin{equation}
\nabla_{{\alpha}}R^{\alpha}\mathstrut_{0,{\lambda}\mu}=-\nabla_{\mu}\rho_{{%
\lambda0}}+\nabla_{{\lambda}}\rho_{\mu0}.
\end{equation}
We call these equations Bianchi quasi-constraints.

\subsubsection{Bianchi evolution system}

The remaining Bianchi equations can be written, as in the case of a Cauchy
adapted frame, as a FOS (first order symmetric) system for two pairs of
`electric' and `magnetic' 2-tensors. This system cannot be said to be
hyperbolic in the usual sense: The principal matrix $M^{0}$ is the unit
matrix hence positive definite but the operators $\partial /\partial t$
appears also in the matrices $M^{i}.$ We say that the system is a quasi-FOSH
system. It is a usual FOSH system with $t$ as a time variable if the matrix
of coefficients of $\partial /\partial t$ is positive definite. It can be
proved that this is the case if the metric induced on the $t=\mathrm{constant%
}$ submanifolds, 
\begin{equation}
\bar{g}_{ij}=\sum_{h}a_{i}^{h}a_{j}^{h}-b_{i}b_{j},
\end{equation}
is positive definite and $U>0$.

\subsubsection{Quasi-FOSH system for connection and frame}

When the Riemann tensor is known the identities which express it become
equations for the connection. Some of them do not contain the derivative $%
\partial _{0},$ we call them connection quasi-constraints. Identities
linking connection and frame become first order equations for the frame
coefficients. No equation gives the evolution of $U$. It can be considered
as a gauge variable fixing the time parameter.

\subsection{Vacuum case}

In vacuum we give arbitrarily on the spacetime $\mathcal{V}$ the scalar $U,$
length of the tangent vector $\partial /\partial i$ to the time line,
together with the projection $Y_{i}$ of $\nabla _{e_{0}}e_{0}$ on $e_{i}.$
The quantities $f_{i}^{j}$ being chosen zero, the identities previously
written give, when the Riemann tensor is known, equations with principal
operator the dragging along the time lines of $\omega _{hj}^{i}$ and $%
X_{ij}, $ and when the connection is known equations for the dragging of the
frame coefficients. These equations together with the Bianchi evolution
equations constitute a quasi-FOSH system. This system is a FOSH system with
respect to $t$ as long as $\bar{g}_{ij}$ is positive definite.

\subsection{Perfect fluid}

In the presence of fluid sources one can obtain a quasi-FOSH system for the
gravitational and fluid variables by taking as time lines the flow lines and
proceeding as follows (Friedrich 1998).

\subsubsection{Fluid equations}

The stress energy tensor of a perfect fluid is, 
\begin{equation*}
T_{\alpha\beta}=(\mu+p)u_{\alpha}u_{\beta}+pg_{\alpha\beta}.
\end{equation*}
Then, 
\begin{equation*}
\rho_{\alpha\beta}=(\mu+p)u_{\alpha}u_{\beta}+\frac{1}{2}(\mu-p)g_{\alpha
\beta},
\end{equation*}
and one supposes that the matter energy density $\mu$ is a given function of
the pressure $p.$

The Euler equations of the fluid, which express the generalized conservation
law $\nabla_{\alpha}T^{\alpha\beta}=0$, are equivalent to the equations, 
\begin{equation*}
(\mu+p)u^{\alpha}\nabla_{\alpha}u^{\beta}+(u^{\alpha}u^{\beta}+g^{\alpha
\beta })\partial_{\alpha}p=0,\text{ \ \ with \ \ \ }u^{\alpha}u_{\alpha}=-1,
\end{equation*}
and, 
\begin{equation*}
(\mu+p)\nabla_{\alpha}u^{\alpha}+u^{\alpha}\partial_{\alpha}\mu=0.
\end{equation*}
In our coframe they read, 
\begin{equation*}
(\mu+p)Y_{i}+\partial_{i}p=0,\ \ \ Y_{i}\equiv\omega_{00}^{i},
\end{equation*}
\begin{equation}
\partial_{0}\mu+(\mu+p)X_{i}^{i}=0.
\end{equation}
Using the index $F$ of the fluid defined by, 
\begin{equation*}
F(p)=\int\frac{dp}{\mu(p)+p},
\end{equation*}
we have, 
\begin{equation}
Y_{i}=-\partial_{i}F,
\end{equation}
and, 
\begin{equation}
\mu_{p}^{\prime}\partial_{0}F+X_{i}^{i}=0.
\end{equation}
The commutation relation between Pfaff derivatives and the definitions give
that, 
\begin{equation*}
(\partial_{0}\partial_{i}-\partial_{i}\partial_{0})F=c_{0i}^{\alpha}
\partial_{\alpha}F=Y_{i}\partial_{0}F-X_{i}^{j}\partial_{j}F,
\end{equation*}
and therefore, 
\begin{equation}
\mu_{p}^{\prime}[\partial_{0}Y_{i}+Y_{i}\partial_{0}F+(f_{i}^{j}-X_{i}^{j})%
\partial_{j}F]-\partial_{i}\mu_{p}^{\prime}\partial_{0}F-\partial
_{i}X_{h}^{h}=0.
\end{equation}
The use of previous identities replaces $\partial_{\alpha}F$ by functions of 
$Y,X$ and $p.$ The derivatives $\partial_{i}\mu_{p}^{\prime}$ are functions
of $Y$ and $p$ since. 
\begin{equation*}
\partial_{i}\mu_{p}^{\prime}=\mu^{\prime}_{p^{2}}\partial_{i}p.
\end{equation*}
Following H. Friedrich, we replace $\partial_{i}X_{h}^{h}$ by its expression
deduced from the equation, 
\begin{equation*}
R_{i0}\equiv\tilde{\nabla}_{h}X_{i}^{\dots\, h}-\tilde{\nabla}
_{i}X_{h}^{h}-Y^{h}(X_{hi}-X_{ih}),
\end{equation*}
and changing names of indices we obtain, 
\begin{equation*}
\mu_{p}^{\prime}\partial_{0}Y_{h}-\tilde{\nabla}
_{j}X_{h}^{j}-Y^{h}(X_{hi}-X_{ih})+Y_{h}\partial_{0}F+
\end{equation*}
\begin{equation}
-X_{h}^{j}\partial_{j}F]+\partial_{h}\mu_{p}^{\prime}\partial_{0}F=0.
\label{10.9}
\end{equation}
Replacing $\tilde{\nabla}_{h}Y_{i}$ by $\tilde{\nabla}
_{i}Y_{h}+c_{hi}^{0}Y_{0}$ where, 
\begin{equation*}
Y_{0}\equiv-\partial_{0}F\equiv-(\mu_{p}^{\prime})^{-1}X_{i}^{i},
\end{equation*}
\begin{equation}
c_{ih}^{0}\equiv\omega_{hi}^{0}-\omega_{ih}^{0}\equiv-\omega_{hi,0}+%
\omega_{ih,0}\equiv X_{ih}-X_{hi},
\end{equation}
we obtain, 
\begin{equation}
\partial_{0}X_{h}^{\dots\, i}-\tilde{\nabla}^{i}Y_{h}+(\mu_{p}^{
\prime})^{-1}X_{i}^{i}(X_{ih}-X_{hi})-Y_{h}Y^{i}+ X_{h}^{\dots\,
j}X_{j}^{\dots\, i}+X_{h}^{\dots\, j}=-R_{h0\dots\, 0}^{\dots\, i}.
\label{10.10}
\end{equation}
The principal operator on the unknowns $Y$ and $X$ in the above equations is
diagonal by blocks and symmetric. The $h$-block reads:

\begin{center}
$\left( 
\begin{tabular}{cccc}
$\mu _{p}^{\prime }\partial _{0}$ & -$\partial _{1}$ & -$\partial _{2}$ & -$%
\partial _{3}$ \\ 
-$\partial _{1}$ & $\partial _{0}$ & 0 & 0 \\ 
-$\partial _{2}$ & 0 & $\partial _{0}$ & 0 \\ 
-$\partial _{3}$ & 0 & 0 & $\partial _{0}$%
\end{tabular}
\right) $
\end{center}

If $\mu_{p}^{\prime}>0$ the matrix $M^{0}$ is positive definite in the 
\textsc{CF}-frame. The system is a quasi-FOSH system for the pairs $%
Y_{h},X_{h}^{j}.$

\begin{remark}
The characteristic determinant associated with the system (\ref{10.9}), (\ref
{10.10}) is, 
\begin{equation*}
\{\xi_{0}^{2}(\mu_{p}^{\prime}\xi_{0}^{2}-\sum_{i=1,2,3}\xi_{i}^{2})\}^{3}.
\end{equation*}
The roots of $\mu_{p}^{\prime}\xi_{0}^{2}-\sum_{i=1,2,3}\xi_{i}^{2}=0$
correspond to sound waves. Their speed is at most 1 (speed of light) if and
only if $\mu_{p}^{\prime}\geq1.$
\end{remark}

\subsubsection{Sources of the Bianchi equations}

In our frame the source tensor $\rho$ reduces to, 
\begin{equation*}
\rho_{00}=\frac{1}{2}(\mu+3p),\text{ \ \ }\rho_{0i}=0,\text{ \ \ }\rho
_{ij}= \frac{1}{2}(\mu-p)\delta_{ij}.
\end{equation*}
We have seen that $\partial_{\alpha}p$ and $\partial_{\alpha}\mu$ are smooth
functions of $p,Y$ and $X.$ The same property holds for $J_{i,0j}$ and J$%
_{i,hk}.$

\subsection{Conclusion}

Assembling the results of the previous subsections we find the following
theorem.

\begin{theorem}
The Einstein equations with source a perfect fluid give a quasi-FOSH system
for the Riemann curvature tensor, the frame and connection coefficients, and
the density of matter, when the flow lines are taken as timelines, $U$ and $%
f_{j}^{i}\equiv\omega_{0j}^{i}$ given arbitrarily.
\end{theorem}

\begin{corollary}
The EEF (Einstein-Euler-Friedrich) system is a FOSH system relatively to $t=%
\mathrm{constant}$ slices as long as the quadratic form, 
\begin{equation}
g_{jh}=\sum_{i=1,2,3}a_{j}^{i}a_{h}^{i}-b_{j}b_{h},
\end{equation}
is positive definite, $U>0$ and $\mu _{p}^{\prime }\geq 1.$
\end{corollary}

\end{document}